\documentclass[sigplan,screen,9pt]{acmart}
\renewcommand\footnotetextcopyrightpermission[1]{}

\usepackage{amsmath,amssymb,amsfonts}
\usepackage{algorithmic}
\usepackage{textcomp}
\usepackage{xcolor}
\usepackage{hyperref}
\usepackage{fancyhdr}
\usepackage{pifont}
\usepackage{makecell}
\usepackage{titlesec}
\usepackage{caption}
\usepackage{float}
\usepackage{tikz}
\usepackage{siunitx}

\usepackage{array}
\usepackage{booktabs}
\usepackage[ruled,vlined,linesnumbered,boxed]{algorithm2e}
\usepackage{stfloats}
\usepackage{threeparttable}
\usepackage{mathrsfs}
\usepackage[mathscr]{euscript}

\usepackage{multirow}
\usepackage{graphicx}
\usepackage{ragged2e} 
\usepackage{flushend}
\usepackage{longtable} 
\usepackage{CJKutf8}
\newcommand{\zdel}[1]{}

 \AtBeginDocument{%
   \providecommand\BibTeX{{%
     \normalfont B\kern-0.5em{\scshape i\kern-0.25em b}\kern-0.8em\TeX}}}

\copyrightyear{
}
\acmYear{
}
\acmDOI{
}

 \acmPrice{
 }
\acmISBN{
}
\begin{document}
\begin{CJK}{UTF8}{gbsn}

\title{SPECRUN: The Danger of Speculative Runahead Execution in Processors}

\author{Chaoqun Shen}
\affiliation{%
 \institution{Hunan University}
 \city{Changsha}
 \country{China}}
\email{shencq@hnu.edu.cn}

\author{Gang Qu}
\affiliation{%
 \institution{University of Maryland, College Park}
 \city{College Park}
 \country{USA}}
\email{gangqu@umd.edu}

\author{Jiliang Zhang}
\authornote{Corresponding authors}
\affiliation{%
 \institution{Hunan University}
 \city{Changsha}
 \country{China}}
\email{zhangjiliang@hnu.edu.cn}

\begin{abstract}

Runahead execution is a continuously evolving microarchitectural technique for processor performance. This paper introduces the first transient execution attack on the runahead execution, called SPECRUN, which exploits the unresolved branch prediction during runahead execution. We show that SPECRUN eliminates the limitation on the number of transient instructions posed by the reorder buffer size, enhancing the exploitability and harmfulness of the attack. We concretely demonstrate a proof-of-concept attack that causes leaking secrets from a victim process, validate the merit of SPECRUN, and design a secure runahead execution scheme. This paper highlights the need to consider the security of potential optimization techniques before implementing them in a processor.
\end{abstract}


\maketitle

\section{Introduction}

Microarchitectural techniques are the crucial foundations for improving modern processors' performance and power efficiency. Despite significant advancements in those metrics, the design of ISA is treated with little security considerations. As demonstrated by Spectre \cite{spectre} and Meltdown \cite{meltdown}, the CPU microarchitecture is susceptible to various vulnerabilities. Attackers may exploit subtle microarchitectural states to extract privileged data \cite{AfterImage}, obtain indirect information \cite{spoiler}, and even access secret keys protected by hardware-enforced primitives such as Intel SGX \cite{MicroScope}. 
Even more alarming is that the slowing of Moore's Law has led to processors with accelerated utilization of more optimization mechanisms to improve performance continually \cite{Pandora}. However, as evidenced in \cite{Pandora,safecracker,value}, some promising techniques also plunge processors into a security crisis.

This work is the first to focus on analyzing a special optimization mechanism, runahead execution \cite{runahead}. It is designed to tolerate long-latency loads that have to be serviced by memory. When memory stalls at the head of the reorder buffer (ROB), the execution enters ``runahead mode'', aiming to speculatively issue multiple independent memory accesses to improve memory-level parallelism (MLP). Since the initial proposal \cite{runahead}, several improvements \cite{filtered,continuous,precise,vector}  have been emerged. Runahead execution has been granted a patent by Intel and has been given the Test of Time Award. Therefore, as one of the most promising techniques for addressing the memory wall problem \cite{vector}, runahead execution is expected to be deployed in future processors. 
However, prior research on the runahead execution has primarily focused on the advancement of performance, with little attention paid to its security.

In this work, we introduce a new attack called SPECRUN, which exploits unresolved branch prediction during runahead execution. We first implement a microarchitectural prototype of runahead execution in a processor simulator. Then, we utilize the designed proof-of-concept (PoC) attack to launch SPECRUN from unprivileged application code, which successfully leaks secret data from a victim's process. 
Moreover, Spectre and its variants \cite{spectre,ret2spec} have an inherent limitation in that the number of transient instructions an attacker can execute depends on the size of ROB. Thus, the attacker cannot execute overflowing gadget instructions due to the limited space. However, SPECRUN eliminates the limitation of ROB size and allows the attacker to execute more gadgets, thereby enhancing the hazard of existing attacks. At last, we propose a secure runahead execution scheme as a potential solution.
 

Our contributions are summarized as follows:
\begin{itemize}

\item \textbf{Security vulnerability.}
We analyze the security of runahead execution for the first time and propose SPECRUN, a novel attack that exploits the interaction between runahead execution and branch prediction. When mixed microarchitectural mechanisms co-working on the same platform, processors may enter an inconsistent state, posing an additional challenge to system security.

\item \textbf{High applicability and unique advantage.} The mixed optimization mechanism is compatible with different runahead execution and branch prediction mechanisms. Additionally, SPECRUN eliminates the limitation posed by the size of ROB on the severity of Spectre-type attacks, enabling more gadgets to be exploited in practical scenarios.

\item \textbf{Secure runahead execution Mechanism.} 
Considering the working principle of runahead execution and the factors contributing to data leakage, we design a secure runahead execution scheme.

\end{itemize}

\vspace{-10pt}
\section{Background}

\subsection{Runahead Execution}
When a long-latency load (called stalling load) causes the instruction window to fill up and halt the pipeline, the processor triggers runahead execution. 
In this process, the stalling load and its dependent instructions are marked invalid using an INV bit. 
These marked instructions the are removed to unblock the pipeline, which allows the processor to execute subsequent independent loads speculatively and expose the MLP. 
Upon the stalling load returns, the processor exits runahead mode and recovers the relevant architectural state. Due to the accurate prefetch requests generated by runahead execution, significant performance benefits are achieved.

The execution of branch predictors warrants particular attention. 
Take the original runahead execution \cite{runahead} as an example. The way branches are predicted and resolved in runahead mode is the same as in normal mode, except for one difference: branches with INV source will never complete resolution.


\subsection{Nested Speculative Execution}

\begin{figure}
\setlength{\abovecaptionskip}{5pt}
\centering
\includegraphics[width=0.8\linewidth]{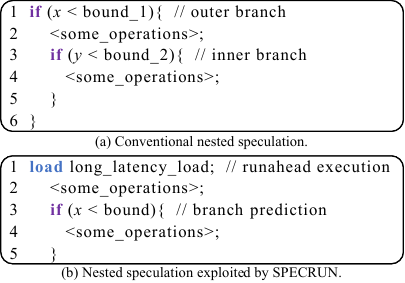}
\caption{The sample code for nested speculative execution.}
 \label{12}
 \vspace{-16 pt}
\end{figure}

Branch instructions constitute a significant portion of the program due to their frequent usage. As a result, the speculative execution path frequently encounters multiple branch instructions. 
The size of the speculative execution window (SEW) can range from tens to hundreds of clock cycles, depending on the triggering event. Modern processors employ nested speculation to avoid front-end pipeline stalls, continuing to predict subsequent branches along the speculative path within the speculative window of the outer branch \cite{branchspec}. In the event that the outer branch is resolved, and a misprediction is detected, the processor discards all instructions executed along the speculative path, including the inner branches. 
As shown in Fig.\ref{12}, previous nested speculation referred to nested branches, while SPECRUN utilizes a new form of nested speculation by combining runahead execution and branch prediction.

\section{Threat Model}


We consider a victim process and an attacker process, where the victim contains secret information that the attacker does not have direct access to but attempts to disclose. Our proposed SPECRUN is a Spectre-type attack, and we adopt the following common assumptions for such attacks:

\begin{itemize}
\item The attacker is co-located with the victim on the same  physical core, but she can only control one process on the target system with normal user privileges.

\item The attacker cannot directly modify or access the victim’s memory, but she has information on the gadget address and the target address to divulge.

\item The attacker can train the target branch through her process and thus mislead the victim to perform vulnerable operations. 

\end{itemize}

Since SPECRUN is based on runahead execution, we require the victim's machine to enable the runahead technology. For research and demonstration purposes, on machines without runahead, Multi2Sim simulation can be used.

\section{SPECRUN Attack}

\subsection{Overview}

Runahead execution improves processor performance by prefetching independent long-latency loads. 
Nevertheless, the speculative nature of runahead execution also introduces potential security issues. We propose a new Spectre-type attack called SPECRUN, 
which exploits incorrect branch prediction within runahead execution. 
The attacker aims to force the victim to enter runahead mode and access sensitive data during branch prediction, ultimately leaking data through a cache covert channel. 


\subsection{Attack Steps}
\begin{figure}
\setlength{\abovecaptionskip}{5pt}
\centering
\includegraphics[width=0.83 \linewidth]{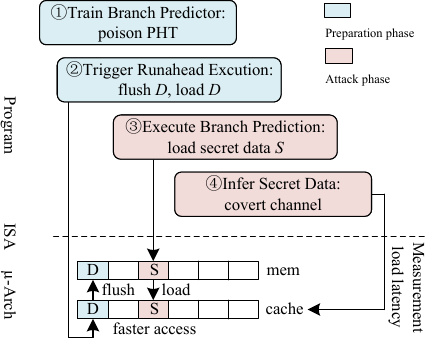}
 \caption{Overview of the workflow of SPECRUN.}
 \label{1}
\vspace{-10 pt}
\end{figure}

Using SpectrePHT \cite{spectre} as an example, Fig.\ref{1} outlines how SPECRUN leaks data through the following four steps. 

\begin{figure}
\setlength{\abovecaptionskip}{5pt}
 \centering
\includegraphics[width=0.9\linewidth]{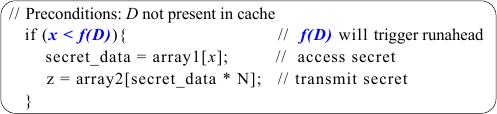}
 \caption{The Code snippet for a possible gadget.}
 \label{2}
 \vspace{-18 pt}
\end{figure}

\textbf{Step 1: Train branch predictor.} 
The attacker runs a vulnerable branch of the victim's code cyclically. The result is that the next time the trained branch is executed, the victim will incorrectly predict the bounds check as true and execute code that should not have been executed, temporarily violating the program semantics. 

\textbf{Step 2: Trigger runahead execution.} 
The operation selected to trigger runahead execution is a long-latency load, which requires the target line to have been evicted from the cache, due to a cache set contention or by employing a co-resident process to execute the $clflush$ instruction. 
Specifically, the latter approach is adopted in this paper, wherein the attacker flushes $D$ and then loads $D$. 

\textbf{Step 3: Execute branch prediction.} The victim performs incorrect speculative execution along the training path and loads a secret-dependent memory block into the cache. 
Notably, the victim's branch predicate depends on the data $D$ manipulated by the attacker. Therefore, the branch has an INV source and cannot completely be resolved during runahead execution. Fig.\ref{2} shows an example of a potential gadget.

\textbf{Step 4: Infer secret data.} The victim leaves traces of secret dependencies in the microarchitecture state. Then the attacker analyses traces to recover the secret data through a covert channel.

\begin{figure*}
\setlength{\abovecaptionskip}{5pt}
 \centering
\includegraphics[width=\linewidth]{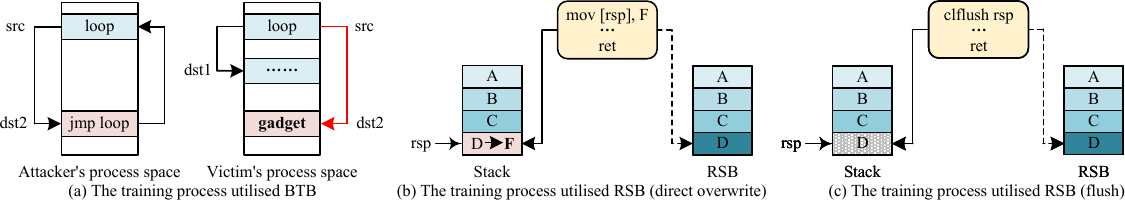}%
 \caption{The mixed optimization mechanism can be applied to Spectre-type attacks. \emph{Gadget} and \emph{F} are associated with stalling load.} 
 \label{16}
 \vspace{-5 pt}
\end{figure*}

\begin{figure}
\setlength{\abovecaptionskip}{5pt}
 \centering
\includegraphics[width=\linewidth]{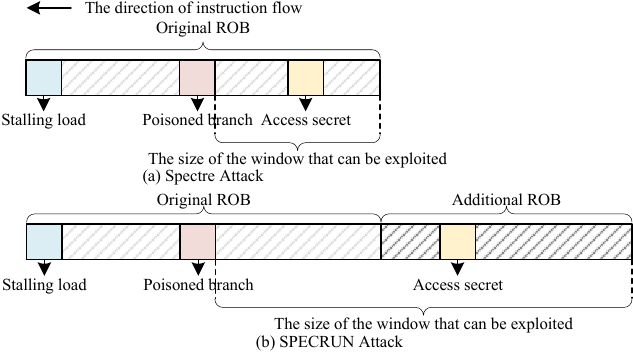}
 \caption{The difference between Spectre and SPECRUN.}
 \label{18}
 \vspace{-20 pt}
\end{figure}

\subsection{Applicability on Runahead Execution Variants}

Precise runahead \cite{precise} and vector runahead \cite{vector} are the latest variants with excellent effects. Branch prediction details are as follows:

\textbf{Precise Runahead.} Only the chains of load instructions (called stall slices) are speculatively executed. Branch instructions are not included in the stall slices because they are not involved in the computation of the load address. However, the front-end relies on the branch predictor to steer the flow of execution in runahead mode, and the branch instructions that reside in the ROB can be resolved and updated as usual. Therefore, if a branch instruction in the ROB has an INV source and is mispredicted, the corresponding stall slice guided by it can be exploited by an attacker to leak data.

\textbf{Vector Runahead.} 
To ensure the consistency of control flow, scalar branches are converted into a predicate mask for the multiple vector lanes, but only the result of the first lane is used to determine the direction of the branch. Moreover, the handling of branch instructions is similar to that in the precise runahead \cite{runahead}. Hence, vector runahead is also susceptible to data leakage risks.

\subsection{Applicability on Spectre Attack Variants}

Once the processor enters the runahead mode, any subsequent type of branch prediction can benefit from the nested speculative execution described above. 
Here are the specific execution details of other Spectre attack variants \cite{spectre,ret2spec}:

\textbf{SpectreBTB.}
As depicted in Fig.\ref{16}(a), the attacker trains the virtual address congruent with $src$ in his own space, which is to design a $loop$ at $src$ to simulate the victim's $loop$, thereby causing the branch originally directed to $dst1$ to instead jump to $dst2$. Similar to SpectrePHT, we associate the gadget address $dst2$ with stalling load to achieve nested speculative execution.

\textbf{SpectreRSB.}  
In the case of a direct overwrite stack, the attacker can enforce incorrect predictions by replacing the return address stored in the stack. Similarly, nested speculative execution can be enabled by associating the polluted value $F$ with a stalling load.
In other cases, the attacker can use $clflush$ to evict the victim's stack address for simplicity, directly achieving nested speculative execution. The details are shown in Fig.\ref{16}(b)-(c).

\subsection{Severity of the Runahead Execution Threat}

Since all changes in the program state during speculation execution must be temporarily stored in internal hardware buffers to enable recovery in case of wrong predictions. Therefore, speculation comes to a halt once any of these buffers become full. In modern Intel CPUs, several buffers can be exhausted \cite{intelsoftware}: ROB, Branch Order Buffer (BOB), Load Buffer (LB), Store Buffer (SB), and Reservation Stations (RS), among others. However, considering factors such as buffer availability (whether the buffer has been recorded and whether the allocated entries will not be reclaimed during speculation) and universality (whether the buffer exists only in specific architectures), the ROB becomes the hardware buffer that limits the maximum number of executable instructions during speculative execution. Specifically, in Intel x86, any speculative path can contain at most as many $\mu$ops as the entries in the ROB.

The essence of runahead execution is to convert the instruction window blocked by long-latency operations into non-blocking windows. This entails removing the stalling load from the instruction window so that subsequent instructions to be pseudo-retired from the ROB and pipeline. In other words, the runahead execution logically increases the size of ROB. Therefore, while the number of transient instructions that can be executed depends on the size of ROB, SPECRUN eliminates this limitation by exploiting the characteristic of runahead execution. SPECRUN presents a significant advantage over previous attacks because its gadgets can accommodate more non-secret-related instructions in addition to the basic instructions related to secret data load and transmission. Overall, as illustrated in Fig.\ref{18}, SPECRUN reduces the requirements for gadgets and possesses a larger SEW.

\section{Experimental Results}

\subsection{Simulation Setup}


\begin{table}
\centering
\renewcommand{\arraystretch}{1.1} 
\vspace{-8pt}
\caption{The basic configuration of the processor.}
\label{table1}
\vspace{-7 pt}
\scalebox{0.83}{ 
\begin{tabular}{|c|c|}
\hline
\textbf{Component}&\textbf{Parameter}\\
\hline
\hline
Core & 2GHz, out-of-order \\
\hline
Processor width & 4-wide fetch/decode/dispatch/commit \\
\hline
Pipeline depth & 6 front-end stages \\
\hline
Branch predictor & two-level adaptive predictor \\
\hline
Functional units & \makecell[c]{4 int add (1 cycle), 2 int mult (2 cycle), \\1 int div (5 cycle), 2 fp add (5 cycle), \\1 fp mult (10 cycle), 1 fp div (15 cycle)}\\
\hline
Register file & 80 int (64 bit), 40 fp (64 bit), 40 xmm (128 bit) \\
\hline
ROB & 256 entries \\
\hline
Queue & i (40), load (40), store (40) \\
\hline
L1 I-cache & 16KB, 4 way, 2 cycle \\
\hline
L1 D-cache & 16KB, 4 way, 2 cycle \\
\hline
L2 cache & 128KB, 8 way, 8 cycle \\
\hline
L3 cache & 4MB, 8 way, 32 cycle \\
\hline
Memory & request-based contention model, 200 cycle \\
\hline
\end{tabular}}
\vspace{-15 pt}
\end{table}

\begin{figure*}
\setlength{\abovecaptionskip}{5pt}
 \centering
\includegraphics[width=\linewidth]{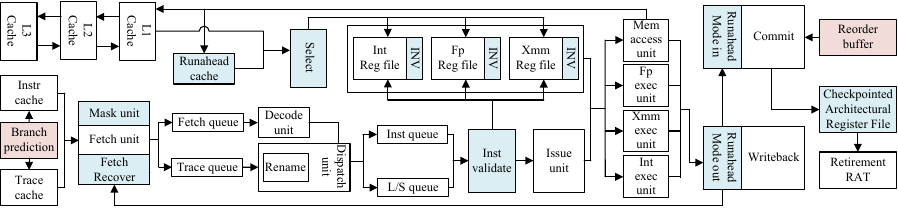}
 \caption{The vulnerable microarchitecture of runahead execution.}
 \label{5}
\vspace{-10 pt}
\end{figure*}

\begin{figure}
\setlength{\abovecaptionskip}{5pt}
\centering
\includegraphics[width=\linewidth]{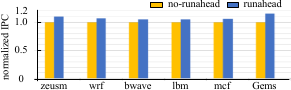}
 \caption{Standardized performance (IPC) comparison.}
 \label{11}
 \vspace{-15 pt}
\end{figure}

We use a runahead execution processor implemented based on the Multi2sim \cite{multi2sim} as the attack object. The detailed configuration is listed in Table \ref{table1}. The corresponding pipeline structure is shown in Fig.\ref{5}. The additional hardware units highlighted in blue are intended to enable the processor to enter, execute and exit the runahead mode at the appropriate times. The components highlighted in red represent the key points of vulnerability exploited by SPECRUN. For more details, please refer to \cite{runahead}.
To evaluate the effectiveness of our approach, we employed SPEC2006 as the benchmark for testing purposes. 
Fig.\ref{11} shows the IPC for machines with and without runahead execution. The results show that we implemented runahead execution and brought an average performance improvement of 11\%. 
Additionally, due to the absence of specific functions in Multi2Sim required to support attack implementations, we need to add two new functions: loading data into the cache and adding a cache flush instruction.

\subsection{Attack Results} 

Fig.\ref{3} presents a PoC example of the SPECRUN, which accesses unauthorised memory by leveraging branch mispredictions in the victim domain during runahead mode.
The victim function receives an integer $x$ from an untrusted source. When $x$ is less than $array1\_size$, the victim accesses array1 with a malicious offset and uses the obtained value as an index for array2 (lines 3-7). To read secret data, we calculate the target address -- the base address of array1 and then call the attacker function with a malicious index ($idx$). 
The attacker function first poisons a branch in the victim's domain (lines 8-12). It then triggers runahead execution, causing the victim to enter runahead mode and run in nested speculative execution (lines 13-16). Finally, the attacker infers the secret data (array1[$idx$]) through the cache covert channel (lines 17-22). Note that the $array1\_size$ in line 3 is associated with the stalling load to allow the victim to enter runahead mode under the attacker's control. 

\begin{figure}
\setlength{\abovecaptionskip}{5pt}
\centering
\includegraphics[width=\linewidth]{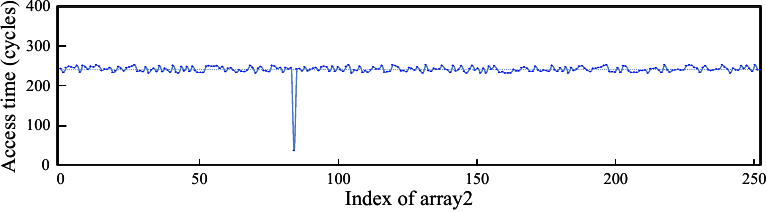}
\caption{The probe array access time after executing SPECRUN.}
\label{7}
\vspace{-15 pt}
\end{figure}

Specifically, the secret data is the array index, and the attacker can infer the data loaded by the victim during execution by measuring the access time of the probe array. If the measured latency is lower than a certain threshold, we can find out that the corresponding index is speculatively loaded. Fig.\ref{7} shows the time the attacker takes to access each value in array2 in turn. It exhibits a significant drop at index 86, indicating that the secret data accessed by the victim is 86. The experiment emphasizes the danger of data leakage due to branch misprediction during runahead execution in the victim domain.

\subsection{The Advantage of SPECRUN}

\begin{figure}
\setlength{\abovecaptionskip}{5pt}
\centering
\includegraphics[width=\linewidth]{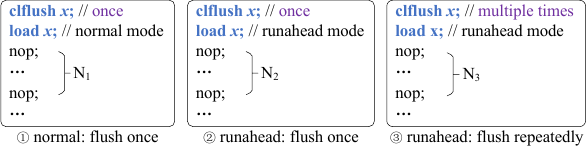}
\caption{The code snippets for measuring the ROB size.}
\label{17}
\vspace{-10 pt}
\end{figure}

By implementing runahead execution, the processor can increase the number of ROB entries in a logical manner, thereby breaking through the limitations imposed by the instruction window. 
Moreover, while the speculative depth of runahead execution is bounded by the available backend resources such as the Issue Queue and Physical Register File \cite{vector}, research detailed in \cite{precise} has demonstrated that the processor possesses sufficient backend resources to sustain speculative execution in runahead mode. Therefore, we can simply use $nop$ instructions, which do not consume backend resources, to validate the effectiveness of increasing the number of ROB entries through runahead execution. As shown in Fig.\ref{17}, we measure the available ROB entries in three execution scenarios, including two cases with extended SEW:

\begin{figure}
\setlength{\abovecaptionskip}{5pt}
\centering\includegraphics[width=0.93\linewidth]{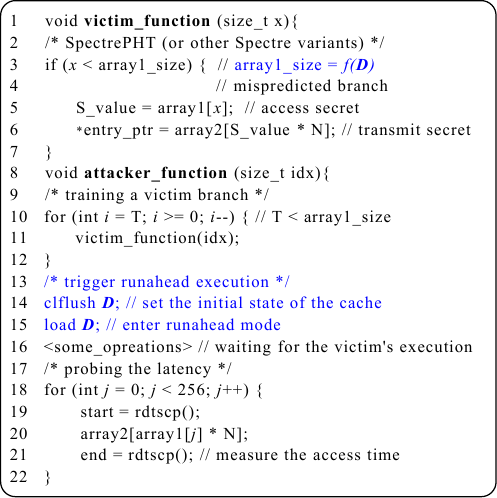}
\caption{The code snippet for SPECRUN.}
 \label{3}
 \vspace{-10 pt}
\end{figure}

\ding{192} Without the runahead execution mechanism, the processor flushes the data ``$x$'' once.

\ding{193} With the runahead execution mechanism, the processor flushes the data ``$x$'' once.

\ding{194} With the runahead execution mechanism, the processor flushes the data ``$x$'' multiple times. The condition that triggers runahead execution is set as a data cache miss, which theoretically allows an attacker to prolong the duration of the processor being in runahead mode by flushing the data multiple times. However, due to thread scheduling, the attacker must wait until all instructions in the ROB have retired before immediately flushing $x$ and repeating this process. And other processes cannot preempt the attacker in the meantime.
Therefore, in real-world execution scenarios, case \ding{194} is a probabilistic event. Our purpose is to point out that there is a possibility for further increasing the size of SEW. 

\begin{figure}
\setlength{\abovecaptionskip}{5pt}
\centering
\includegraphics[width=\linewidth]{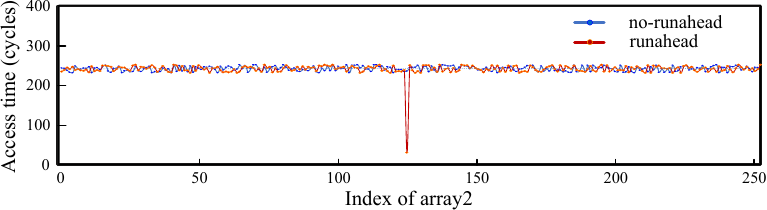}
\caption{The access time of the probe array on a no-runahead machine and a runahead machine.}
\label{22}
\vspace{-15 pt}
\end{figure}

Based on our measurements, the respective values for $N_1$, $N_2$, and $N_3$ are 255, 480, and 840, while ROB has 256 entries. Thus, it is evident that the number of executable instructions in runahead mode exceeds the capacity of ROB. To verify this advantage of the SPECRUN, we modify the code shown in Fig.\ref{3} by inserting multiple $nop$ instructions in line 4 to ensure that the access instruction for secret data (line 5) appears outside the original ROB entries. We then execute the modified code on a no-runahead machine and a runahead machine, respectively. As shown in Fig.\ref{22}, the experimental results exhibit that no significant latency drop is observed on the no-runahead machine, meaning the absence of data leakage. 
However, on the runahead machine, data leakage can still be observed at index 127. Hence, this proves that runahead-based speculative execution attacks can pose more significant harm, introducing the risk of data leakage to initially secure code.

\section{A Secure Runahead Execution Mechanism}

While the processor enters runahead mode, it speculatively executes subsequent instructions and loads data into the cache, creating a natural attack surface. We introduce a Speculative Load cache (SL cache) to address this vulnerability to store data loaded during runahead execution. Functioning as an ``L0 cache'', the processor first accesses the SL cache upon exiting runahead mode. If a cache miss occurs, the data is searched from the L1 cache as usual. 
If the data exists in the SL cache, additional checks are required to determine whether it can be loaded into the L1 cache. 
To this end, we utilize the concept of Unsafe Speculative Load (USL) from InvisiSpec \cite{invisispec} to guide data placement. 
However, since branches in SPECRUN cannot be resolved, we restrict the definition of USL to the loads related to secret.

To be more specific, we employ taint tracking to identify the USLs during runahead execution. We introduce the tags $B_{tag}$ and $IS$ in the SL Cache. The variable $B_n$ is defined as the $n$th branch, and $B_{tag}$ = $B_{n,m}$ indicates that the load is the $m$th USL (i.e., taint-related load) within the scope of the $n$th branch. For the untainted load within the scope of $B_n$, $B_{tag}$ is represented as $B_{n,0}$, while for the load outside the branch scope, $B_{tag}$ is 0. In addition, the $IS$ indicates whether the output of an instruction is associated with the taint of $B_n$. If the instruction's input depends on tainted data within the scope of $B_n$, $IS$ is set to $B_n$; otherwise, $IS$ is set to 0, meaning the load is safe. 
In this scheme, $B_{tag}$ determines the security of the loads within the branch scope, while $IS$ marks all potential USLs related to tainted data. 
However, identifying all tainted loads faces two challenges. First, nested branches need to be identified. 
Second, the taint-related loads outside the branch scope must be marked. 
The following is the specific solution to address these challenges. 

\begin{figure}
\setlength{\abovecaptionskip}{5pt}
 \centering
\includegraphics[width=\linewidth]{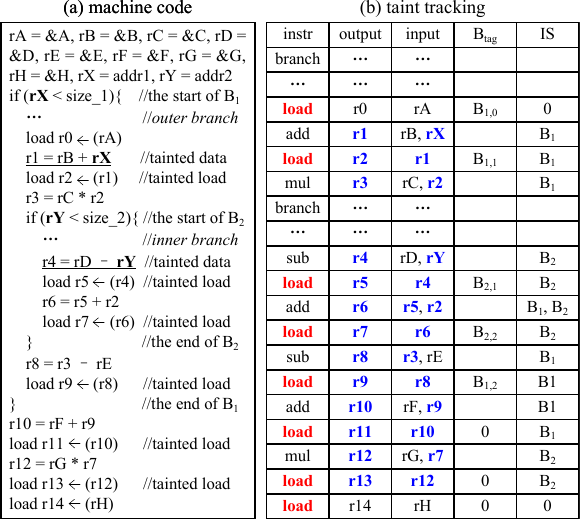}
 \caption{The process of marking $B_{tag}$ and $IS$. The processor encounters $B_2$ during the matching process for $B_{1e}$, indicating that branches $B_1$ and $B_2$ are nested. 
Load instructions are \textcolor{red}{ \textbf{red}} and tainted registers are \textcolor{blue}{\textbf{blue}}.
 }
 \label{21}
\vspace{-17 pt}
\end{figure}

The processor utilizes $B_{ns}$ and $B_{ne}$ to represent the start and end addresses of branch $B_n$ based on the compiled code. If the processor encounters a new branch, such as $B_y$ before matching $B_{xe}$, it means that $B_y$ is an inner branch of $B_x$. Consequently, the matching order is changed to first match the end address of the inner branch, followed by that of the outer branch. When the processor needs to match $B_{ne}$, the loads during this period are marked as $B_{n,M}$ (M $\in$ [0,m]). With these guidelines, we can effectively record the nested relationships of branches and mark the loads with $B_{n,m}$. Meanwhile, we determine the $IS$ based on the tainted data propagation and the instruction inputs' corresponding branch information. 
The process of marking $B_{tag}$ and $IS$ is shown in Fig.\ref{21}.

Additionally, to mitigate the negative performance impact of unnecessary SL cache accesses, we add a counter $C$ to record the number of data entries residing in the SL cache. Once $C$ is reduced to 0, the processor ceases accessing the SL cache and instead loads data through the regular memory subsystem. The process of loading data by the processor after exiting the runahead mode is outlined in Algorithm \ref{Pro}. Concretely, we choose different methods to reduce the value of $C$ depending on the data type:

1) If the data retrieved from the SL cache is considered safe ($IS$ = 0), it is placed in the L1 cache, removed from the SL cache, and the value of $C$ is decremented by 1.

2) In the case of USL ($IS$ $\neq$ 0), the processor waits for branch resolution. 
If the branch prediction is incorrect, the pre-stored nested branch information is utilized to determine which loads to remove. 
Specifically, the processor uses $IS$ to delete the loads associated with the incorrectly executed branch itself and its inner branches. After that, the value of $C$ is reduced based on the number of deleted entries. Otherwise, the execution is the same as above.


In addition to designing a secure runahead execution scheme, we can also consider imposing restrictions on existing mechanisms. For example, we can nullify the impact of branches on instruction execution within the runahead interval. Once a branch predicate is identified as associated with a stalling load, the branch is skipped rather than unresolved. 
In summary, before deploying runahead execution, it is crucial to balance security and performance carefully. While effective,  the countermeasures may lead to increased overhead and degraded performance to some extent.


\setlength{\textfloatsep}{3pt}
\begin{algorithm}[tbp]
\algsetup{linenosize=\normalsize} 
\small
\renewcommand{\algorithmcfname}{Algorithm}
\DontPrintSemicolon
\caption{The load process after exiting runahead mode
}
\label{Pro}
\KwIn{$S[ ]$, $D_{SL}$, $d$, $B_{n}$, $B_{tag}$, $IS$, $C$}
$S[ ]$: An array that stores the branches judged to be correctly predicted\;
$D_{SL}$: The data fetched from the SL cache\;
$d$: The number of entries in the SL cache that need to be deleted due to USL\;
processor encounters a load instruction; \;
\If {$C$ $\neq$ 0}{ 
    load data $D_{SL}$ from the SL cache;\;
    \If {$B_{tag}$ $\neq$ 0}{
        extract the $B_n$ information;\;
        \If {$B_n$ is not in $S[ ]$}{
            wait the resolution of $B_n$;\;
            \If {the prediction of $B_n$ is correct}{
                $S[ ] \leftarrow B_n$;\;
                $D_{SL}$ is loaded to L1 cache;\;
                $C$ = $C$ - 1;\;
            }
            \Else {
                use $IS$ to delete entries related to $B_n$;\;
                $C$ = $C$ - $d$;\;
            }
        }
        \Else {
            $D_{SL}$ is loaded to L1 cache;\;
            $C$ = $C$ - 1;\;
        }
    }
    \Else {
        $D_{SL}$ is loaded to L1 cache;\;
        $C$ = $C$ - 1;\;
    }
}
\Else {
    loads data from the regular cache;\;
    switch execution paths so that subsequent loads before the next round of runahead execution no longer access the SL cache;
}
\end{algorithm}

\vspace{-15pt}
\section{Related Work}
A large number of attack variants have been proposed based on optimization mechanisms of commercial processors such as out-of-order execution, branch prediction, store-to-load forwarding, memory disambiguation, speculative dependency checking, prefetching, fault handling, microcode assist, and others. Microarchitectural vulnerabilities have sparked widespread concern, as they leak arbitrary confidential information in a reliable and fine-grained manner. What is even more concerning is that as Moore's Law slows down, future processors may adopt various technologies to enhance performance, which may also introduce unknown security issues. To address this impending security crisis, a trend of the current studies is to perform prospective security analysis on optimization mechanisms that may be deployed in the future. Several recent works have focused on this topic.

\cite{Pandora} proposes a conceptual framework to study seven classes of microarchitectural optimization with security implications, ranging from value locality to compressibility to prediction to prefetching, then provides a PoC implementation of silent stores and data memory-dependent. \cite{safecracker} offers the first security analysis of cache compression and introduces a new concealing side channel, as simply storing data transmits information about it.
\cite{value} proposes a new security attack on value predictor. Furthermore, the authors developed a model for analyzing value predictor attacks, based on which they found 12 attack variants.

Our proposed SPECRUN leverages the potential of runahead execution, an optimization mechanism aimed at mitigating the performance impact of long-latency accesses.
SPECRUN has simple trigger conditions and high applicability compared to previous works. What sets it apart is the utilization of a mixture of optimization mechanisms, i.e., nested branch prediction within runahead execution. 
This allows SPECRUN to eliminate the limitation on the number of transient instructions imposed by the ROB size, leading to more severe security threats.

\section{Conclusion}
Runahead execution is a promising microarchitectural technique designed to mitigate the memory wall. However, its security has yet to be evaluated prior to this work. As demonstrated in this study, runahead execution is also threatened by transient execution attacks. We propose a novel attack named SPECRUN, which takes advantage of the microarchitectural design of unresolved branch prediction in runahead processors and eliminates the limitation on the number of transient instructions. In light of this, we also design a secure runahead execution scheme based on invisible and tainted tracking. 
We hope that this research will prompt careful consideration of security before implementing optimization mechanisms in future commercial processors.


\end{CJK}
\end{document}